# Speech User Interface for Information Retrieval

Urmila Shrawankar
Dept. of Information Technology
Govt. Polytechnic Institute, Nagpur
Sadar, Nagpur – 440001 (INDIA)
urmilas@rediffmail.com
Cell : +919422803996

Anjali Mahajan
Dept. of Computer Sci & Engg.
G.H. Raisoni College of Engg.,
Hingna, Nagpur – (INDIA)
armahajan@rediffmail.com

**Abstract:** Along with the rapid development of information technology, the amount of information generated at a given time far exceeds human's ability to organize, search, and manipulate without the help of automatic systems. Now a days so many tools and techniques are available for storage and retrieval of information. User uses interface to interact with these techniques, mostly text user interface (TUI) or graphical user interface (GUI). Here, I am trying to introduce a new interface i.e. *speech* for information retrieval.

The goal of this project is to develop a speech interface that can search and read the required information from the database effectively, efficiently and more friendly.

This tool will be highly useful to blind people, they will able to demand the information to the computer by giving voice command/s (keyword) through microphone and listen the required information using speaker or headphones.

## 1 Introduction

Digital information organization refers to methods of rendering large amounts of information into digital form so it can be stored, retrieved and manipulated by computer. An example of digital information organization is the digital library.

Information Retrieval (IR) is a discipline of studying theories, models, and techniques that deal with the representation, storage, organization, and retrieval of information items so that they can be useful to humans.

The rapid rise in computer and Internet use has resulted in the creation of vast quantities of digital information being created, retrieved and transmitted.

This project will help to store the information in the database such a way that the information can be retrieved by speech (voice command inputted through microphone) on demand. The retrieved information can be view on the computer screen and listen through speakers connected to the computer.

The project model is as below:

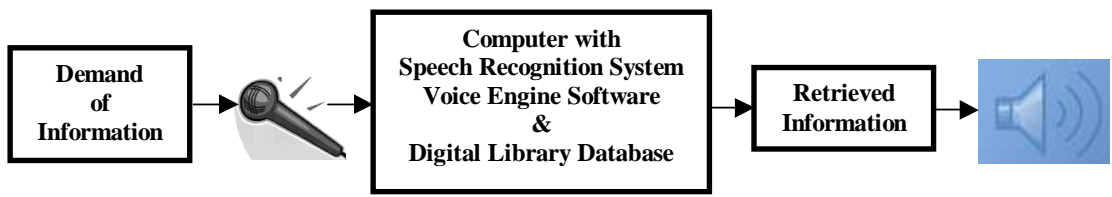

*<u>Project Model</u>*

The idea behind this project is develop a database with keywords and implement Voice Engine.
In the database along with the usual fields, one more field is added that is *keyword.*
This keyword (Search word) field should be properly indexed for search. The keyword field is also essential in case of picture objects, along with picture information.
A Voice Engine is developed to train and recognise the word. It takes care of accepting speech input, extracting the features from analog speech signal and converts them into parametric form (speech-to-text) for further processing. The Linear Predictive Coding (LPC) model is implemented for speech-to-text conversion. Voice Engine trains the machine for sample words using Artificial Neural Network (ANN) techniques. The system further developed for recognise the word using Hidden Markov Model (HMM). Finally retrieved information gets converted into text-to-speech format.
The user has to speak out the keyword through microphone connected to the computer. Computer will pass the keyword to the Voice Engine. Using recognised word, a query (SQL) will generated and pass to database. The searched information will retrieve from database and displays on computer screen and further that text will convert into speech so that user can listen through speaker or headphones. In case of picture, picture will display on the screen and related picture information will convert into speech.
In case of failure in recognising and retrieving information appropriate messages flashes on the error window and can be hear through speaker.

## 2 Information Storage and Retrieval:

Computer based automatic digital library development project is based on followings factors:

**Storage:** A digital library's storage system must be capable of storing a large amount of data in a variety of formats and accessing this data as quickly as possible. Such as text, documents, pictures, audio and video etc.
This project considers the SQL server to store the database. Information stores in form of various inter-related tables with key field *keyword*.

**User interface:** The user interface, perhaps the most important digital library component, must incorporate a wide variety of techniques to afford rich interaction between users and the information they seek. For computer workstations, apart from text and graphical user interfaces this project introduces a *speech user interface* that will be very helpful to blind, old-aged and other physically disabled people.

**Speech user Interface over the network**: This project uses the speech user interface on client side only, query hits to the database server as the regular text query. The Voice Engine is employed on the client side therefore; either in the client-sever architecture or intranet and internet environment working will be as the regular style. In the network environment no need to use special protocol to understand speech command.

**Information retrieval:** The concepts underlying information retrieval were conceived long before computers and information systems were employed to store library materials. In the digital library domain, there are a variety of information-retrieval techniques, including metadata searching, full-text document searching, and content searching for other data types. This project mostly focuses on content searching with the help of keyword. The success of information retrieval can be measured in terms of the percentage of relevant and extraneous information retrieved. It is difficult to pinpoint quantitatively the effectiveness of information retrieval; only an individual user can determine what is truly useful. Techniques to improve retrieval effectiveness include pre-processing documents to extract additional metadata before storing them in a document database.

# 3 Speech Recognition System

Speech Recognition is a technology, which allows control of machines by voice in the form of isolated or connected word sequences. It involves the recognition and understanding of spoken language by machine.

Speech Recognition is fundamentally a pattern classification task. The objective is to take an input pattern, the speech signal and classify it as a sequence of stored patterns that have precisely been defined. These stored patterns may be made of units, which we call *phonemes*.

If these patterns were invariant and unchanging, the problem would be trivial; simply compare sequences of features with the stored patterns, and find exact matches when they occur. But the fundamental difficulty of speech recognition is that the speech signal is highly variable due to different speakers, different speaking rates, different contents and different acoustic conditions.

Based on the difference in the way, words are pronounced, there are following three standard modes of speaking to a machine, namely;
- Isolated Word Recognition
- Connected Word Recognition
- Continuous Speech Recognition

Degree of speaker dependence is based on whether the recognition system is:
- Speaker Dependent
- Speaker Independent

This project uses *Isolated Word*, *Speaker Independent* Speech Recognition System.

# 4 Feature Extractions and Feature Matching

Feature extraction is the process that extracts a small amount of data from the voice that can later be used to represent each word. Feature matching involves the actual procedure to identify the new word by comparing extracted features from stored samples.

All speaker recognition systems have to serve two distinguishes phases. The first one is referred to the enrollment sessions or training phase while the other is referred to as the operation sessions or testing phase. In the *training phase*, for each keyword form different speakers have to provide samples of their speech so that the system can train a reference model. During the testing phase, the input speech is matched with stored reference model(s) and recognition decision is made.

# 5 Approaches To Automatic Speech Recognition By Machine

The Speech recognition by the machine, whereby the machine attempts to decode the speech signal in a sequential manner based on the observed acoustic features of the signal and the known relations between acoustic features and phonetic symbols.

There are several proposed approaches to automatic speech recognition by machine with the goal of providing some understanding as to the essential of each proposed method and the basic strengths and weaknesses of each approach.

Broadly speaking, there are three approaches to speech recognition, this project based on Artificial Intelligence Approach. Others are namely,

- The Acoustic-Phonetic Approach
- The Pattern Recognition Approach
- The Artificial Intelligence Approach

# 6 Signal Processing And Analysis Methods For Speech Recognition

Signal processing is the process of extracting relevant information from the speech signal in an efficient, robust manner.
A speech recognition system comprises a collection of algorithms drawn from a wide variety of disciplines. All recognition systems are the signal processing front end, which converts the speech waveform to some type of parametric representation for further analysis and processing.

# 7 Speech Feature Extraction

The purpose of this module is to convert the speech waveform to some type of parametric representation for further analysis and processing. This is often referred as the *signal-processing front end*. There are many models are available for analysis like,

- Linear Predictive Coding (LPC) Model
- Mel-Frequency Cepstrum Coefficients (MFCC) Model
- Bank-of-Filters
- Vector Quantization

## Linear Predictive Coding (LPC) Model

This project uses Linear Predictive Coding (**LPC**) model, it is one of the most powerful speech analysis techniques and is a useful method for encoding quality speech at a low bit rate. It provides accurate estimates of speech parameters and efficient for computations.
LPC system is used to determine the formants from the speech signal. The basic solution is a difference equation, which expresses each sample of the signal as a linear combination of previous samples. Such an equation is called a linear predictor, which is why this is called Linear Predictive Coding.

# 8 Training

## Artificial Neural Networks (ANN)
Neural networks are often used as a powerful discriminating classifier for tasks in automatic speech recognition. They have several advantages over parametric classifiers. However there are disadvantages in terms of amount of training data required, and length of training time. Some neural network architectures are:
- Feedforward Perceptrons Trained With BackPropagation
- Radial Basis Function (RBF) Networks
- Learning Vector Quantization (LVQ) Networks

Neural networks training with Feedforward back propagation is most popular model and it is used in speech recognition systems. However, the length of time required to train the networks can present problems, particularly when investigating a variety of feature sets to represent speech data.

## Backpropagation Model
Backpropagation uses a gradient descent approach to minimize output error in a feed-forward network. The algorithm involves presenting an input vector, comparing the network output to the desired output for that vector, and updating each weight by an amount corresponding to the derivative of the error with respect to that weight times some learning rate, and adjusted by a "momentum" factor.

### Hidden Markov Model (HMM)

In the context of statistical methods for speech recognition, hidden Markov models (HMM) have become a well known and widely used statistical approach to characterizing the spectral properties of frames of speech. As a stochastic modeling tool, HMMs have an advantage of providing a natural and highly reliable way of recognizing speech for a wide variety of applications. Since the HMM also integrates well into systems incorporating information about both acoustics and syntax, it is currently the predominant approach for speech recognition. Hidden Markov Models HMMs are "doubly stochastic process" in which the observed data are viewed as the result of having passed the true (hidden) process through a function that produces the second process (observed). The hidden process consists of a collection of states connected by transitions. Each transition is described by two sets of probabilities:
• A **transition probability**, which provides the probability of making a transition from one state to another.
• An **output probability** density function, which defines the conditional probability of observing
Using the combination of following algorithms the HMM is implemented.

- Forward - Backward algorithm
- Viterbi algorithm
- Baum -Welch algorithm

## 9 Recognition

For continuous speech recognition applied to large-vocabulary tasks, the search algorithm needs to apply all available acoustic and linguistic knowledge to maximize recognition accuracy. In order to integrate the use of all the lexical, linguistic, and acoustic sources of knowledge, this approach uses the Viterbi algorithm as a fast-match algorithm, and a detailed re-scoring approach to the N-best hypotheses to produce the final recognition output.
The decoder is designed to exploit all available acoustic and linguistic knowledge in several search phases. Initially, a Viterbi - Baum search produces a single recognition hypothesis as well as a word lattice that includes word segmentations and acoustic scores. The word lattice is transformed into a directed a cyclic graph (DAG). With DAGs, it is possible to use other possibly larger language models and perform a much quicker search for the best hypothesis.

## 10 Conclusions

Speech understanding by the machine and interacting with the human like human-to-human will be the real interface for human-to-machine interaction. To develop such interface training the machine by providing artificial intelligence is necessary. If machine gets train property we get good result at recognition phase.
Some adverse condition like noise, affects the performance. Noise can be reduce by situating machine at noise free environment.
The Speech interface for information retrieval in digital library system will be the boon for blinds, physically challenged people and aged people.
Such applications can be further developed for different National and Regional languages.

- *Cole, Vuuren* "**Perceptive animated interface: First steps toward a new paradigm for human –computer Interation**"
- *Adrid, Barjaktarevic, Ozum*, "**Automatic Speech Recognition for Isolated Words**"

**c) Journals**
- *Lawrence Rabiner*, "**A Tutorial on Hidden Markov Models and Selected Applications in Speech Recognition**", Roceedings of The IEEE, Vol. 77, No2., Feb-1989
- *C.J. Legetter, P.C. Woodland*, "**Maximum likelihood linear regression for speaker adaptation of continuous density hidden Markov models**," Computer Speech & Language, Vol. 9, pp. 171-185, 1995.

**d) Reports**
- *Pellom, B*. "**Sonic: The University of Colorado Continuous Speech Recognizer**", Technical Report TR-CSLR-2001-01, CSLR, University of Colorado, March 2001.
- *M. J. F. Gales*, "**Maximum Likelihood Linear Transformations for HMM-Based Speech Recognition**," Technical Report CUED/F-INFENG/TR 291, Cambridge University, May 1997

**e) Wrbsite**
- Exploratory Development Of **Informaiton Retrieval** Systems At Stic **... Digital Library** Initiative At The Nanyang Techno Logical University**...** Web.Simmons.Edu/~Chen/Nit/Procnit99.Html
- Class Work: Understanding **Informaiton Retrieval** Basics **...** Halvorsen, Per-Kristian ; Robertson, George G. (1995) Rich Interaction In The **Digital Library**. **...** www.Ischool.Utexas.Edu/~I385d/Schedule.Html -
- New **Information** Technology 2001 Conference. Global **Digital Library** Development In The New **...** Sigir-96 Workshop On Networked **Information Retrieval**. **...** Bengross.Com/Dl/

**f) Paper in published proceedings**
- *Umit H. Yapanel, John H.L.Hansen*, "**A new perspective on Feature Extraction for Robust Invehicle Speech Recognition**" Proceedings of Eurospeech'03, Geneva, Sept. 2003.